# Magnetically-driven Structural Phase Transition in Hexamethylbenzene

Fei Yen,[1,2*]

[1]School of Sciences, Harbin Institute of Technology (Shenzhen), University Town, Shenzhen, Guangdong 518055, P. R. China
[2]State Key Laboratory on Tunable Laser Technology, Ministry of Industry and Information Technology Key Laboratory of Micro-Nano Optoelectronic Information System, Harbin Institute of Technology (Shenzhen), University Town, Shenzhen, Guangdong 518055, P. R. China

**Correspondence:** *fyen@hit.edu.cn, fyen18@hotmail.com

**Abstract:** The methyl groups in hexamethylbenzene $C_6(CH_3)_6$ become magnetically ordered at the molecular level below 118 K. This is also near the temperature at which the system structurally transitions from triclinic to a unique near-cubic phase. High-precision measurements of the near-static dielectric constant reveal that the structural phase transition is actually comprised of four successive transformations upon cooling at $T_1$=110.7 K, $T_2$=109.5 K, $T_3$=109.1 K and $T_4$=107.8 K. In contrast to warming, only two transitions occur at $T'_4$=119.2 K and $T'_1$=120.9 K. The methyl groups in the near-cubic phase become slightly distorted according to existing neutron powder diffraction measurements. Analysis of the $2^6$=64 possible spin orientation configurations of the methyl groups reveal a 20-fold ground state degeneracy presiding in each molecule rendering the system to become highly unstable. From such, it is interpreted that $T_1$ and $T_2$ are temperatures at which the molecules successively lower their symmetry to remove the energy degeneracy which involve methyl group elongation and further out-of-plane tilting. This triggers the system to phase transition into the near-cubic phase which involves a shearing of the molecular planes and partial rotation of the methyl groups at $T_3$ and $T_4$. We interpret the low temperature near-cubic phase to be a manifestation of Jahn-Teller distortions based on energy degeneracies of the orbital motion of protons in each molecule and suggest that the metastable nature of the phase transition originates from the methyl groups requiring a larger amount of energy to order than to disorder. Our findings help explain why many unusual structural phase transitions occur at low temperatures in other molecular crystals possessing periodic motion of protons.

*{This manuscript is the unedited version of the Author's submitted work which was subsequently accepted (December 4th, 2018) for publication in the Journal of Physical Chemistry C after peer review. To access the final edited and published version of this work, please visit [https://pubs.acs.org/doi/10.1021/acs.jpcc.8b09864]. Thank You.}*

**Introduction:**

Hexamethylbenzene $C_6(CH_3)_6$ (HMB) possesses three solid phases, an orthorhombic phase **I**[1] from 383–439 K; a triclinic phase **II**[2,3] from 118–383 K and a near-cubic phase **III** with a unique set of lattice parameters[4] below $T_{C\_H}$=118 K. At room temperature, weak cohesive forces allow for entire molecules to rotate about their $C_6$ axes.[5-7] Below 160 K, the $C_6$ rotations seize,[8] however, the methyl groups continue to rotate freely about their respective $C_3$ axes. The carbon atoms of the methyl groups lie on the axes of rotation so together with the benzene rings, the carbon atoms are treated as being rigid exhibiting no translational motion. As a methyl group completes one revolution, its three protons ($H^+$) at the extremities encircle an area $A$. Such periodic movement of three positive charges at a particular speed equates to a current $I$ from which a magnetic moment $\mu_p=IA$ arises. Only one degree of freedom exists and that is whether the methyl group rotates clockwise or anticlockwise which result in $\mu_p$ pointing radially inward or outward, respectively. Figure 1a shows the $2^6$=64 possible spin configurations of $\mu_p$ that each HMB molecule can possess when the methyl groups rotate freely. Despite that the mass of protons is ~1836 times larger than electrons (therefore possessing a current smaller by the same factor), the area the electrons encircle is ~$10^{10}$ times smaller (the path an electron travels in one revolution while undergoing Larmor precession is ~$2\pi \times 10^{-15}$ m while the distances traveled by the protons here is ~$10^{-10}$ m). Even if the C-H bond electrons, if treated on average to lie midway the carbon atoms and protons, are also taken into consideration to rotate with the protons, the area these electrons encircle is only a quarter that of the protons. It is clear to see that $\mu_p > \mu_e$ in HMB. To our knowledge, all studies on HMB, in fact on all molecular solids that exhibit periodic rotational motion of protons, have only emphasized on the nuclear magnetic moments and electronic structures; it is apparent that the 'protonic configurations' in such compounds should also be investigated as they may play a key role in determining the physical properties of the system.

Recently, magnetic susceptibility measurements on hydrogenous[9] and deuterated hexamethylbenzene (HMB-$d_{18}$)[10] revealed that the methyl groups become magnetically ordered at the molecular level below $T_{C\_H}$. According to numerous theoretical and computational studies the methyl groups in each molecule become 'geared' with alternating spins pointing inward and outward as this is the configuration with the lowest energy.[11-16] A careful inspection of the neutron powder diffraction measurements reveal that the methyl groups become elongated and slanted in the **III** phase[4]. From the supplementary information provided in Ref. 4, the D-C-D (where D is deuterium) and D-C-C angles at different temperatures are shown in Figs. 2a and 2b. One likely scenario is that as the temperature is cooled past $T_{C\_H}$, Jahn-Teller distortions occur as the system needs to lift a 20-fold spatial energy degeneracy of its methyl groups depicted in Fig. 1a. Dielectric constant $\varepsilon'(T)$ measurements are extremely sensitive to geometric changes at the molecular level[17-19] and can be measured nearly continuously with respect to temperature. In some cases, even magnetic ordering and spin reorientations may be detected when there exists a coupling between the order parameters.[20-22] For this reason, we carried out high-precision measurements of $\varepsilon'(T)$ and find that the molecular distortions are comprised of a series of four steps upon cooling. This is in contrast to warming where the reverse process occurs in nearly one single step. These results are employed to discuss the origin of metastability between the **II** and **III** phases and conjecture that similar magnetically driven structural transitions occur in many other molecular solids.

**Methods:**

The samples were acquired from Sigma Aldrich synthesized via sublimation and >99.9% in purity. The crystals were thin, transparent and plate-like with dimensions of roughly 1.5 mm x 2.5 mm and 200 microns thick. Copper wire 25 μm in diameter were attached to both sides of the crystal by silver paint forming a parallel plate capacitor. The dielectric constant was derived from the measured capacitance of the two electrodes by an Andeen-Hagerling ultra-high precision capacitance bridge at 10 kHz with a bias of 15 V. The temperature was controlled to change at the rate of 0.5-1 K/min by the cryostat of a PPMS (Physical Property Measurement System) unit manufactured by Quantum Design, Inc., U.S.A.

**Results:**

Figure 3 shows the temperature dependence of the dielectric constant $\varepsilon'(T)$ at 10 kHz of HMB measured along the direction perpendicular to the molecular planes. Upon cooling, a sharp change in slope discontinuity occurs at $T_1=110.7$ K. At $T_2=109.5$, $\varepsilon'(T)$ exhibits a sharp minimum and its slope becomes negative. Then a sharp peak occurs at $T_3=109.1$ K. Finally, $\varepsilon'(T)$ exhibits one last change in slope at $T_4=107.8$. Upon warming, no discontinuities occur up until $T'_4=119.2$ where the degree of the slope of $\varepsilon'(T)$ changes to almost the same as in the case when $T_4 < T < T_3$. This trend is only briefly sustained up to $T'_1=120.9$ K where $\varepsilon'(T)$ retraces its original path. In between $T'_4$ and $T'_1$ also lies an extremely small peak in what appears to be the $T'_3$ and $T'_2$ transition temperatures occurring somewhere near 120.0 K. $T_1$ and $T'_4$ coincide with the step-up and step down anomalies in the magnetic susceptibility, respectively, reported previously in Refs. 9 and 10.

**Discussion:**

The four observed anomalies where $\varepsilon'(T)$ abruptly changes its slope can only be attributed to geometric transformations. According to neutron powder diffraction data, the transition from **III** to **II** during warming involves the shearing of the molecular planes along with a partial rotation of the methyl groups[4] which is what $T'_4$ and $T'_1$ represent. Based on our results, it is clearly evident that two additional transformations occur when the system phase transitions from **II** to **III**.

We now attempt to decipher why two additional distortions occur upon cooling. As the system is cooled past $T'_1$ the system wants to order, but the spin configurations in each molecule are random as shown in Fig. 1a. The configurations with the lowest energy are when three arrows point inward and three outward (since neither the spin-in nor spin-out represent a negative value so the lowest attainable value is zero). However, 20 out of the 64 possible configurations meet this criterion. According to the Jahn-Teller theorem,[23] the system must find a way to physically lower this 20-fold spatially degenerate ground state. An elongation of the methyl groups *1* and *4* reduces the degeneracy to 12-fold as this restricts sites *1* and *4* labeled in Fig. 1b to have to oppose each other so the energy remains zero (the value of their magnetic moments are the same and cancel each other out only when their vector sum vanish). Elongation of the methyl groups *3* and *6* to a different value further reduces the degeneracy to 8-fold (Fig. 1c) as this restricts sites *3* and *6* as well as *2* and *5* to have to oppose each other. To further obtain a minimum, methyl groups *1* and *4* become more tilted out-of-plane which cause their nearest adjacent methyl groups at sites *2* and *5* to spin in the opposite direction and reduce the energy degeneracy to 4-fold (Fig. 1d). Lastly, methyl groups *3* and *6* also tilt further out-of-plane to a different angle rendering all adjacent methyl groups to spin along opposite directions (Fig. 1e). It should be noted that the abovementioned distortions may not follow in the order as described. However, after lowering its symmetry by readjusting the H-C-H and H-C-C angles to three unique pair values (which deviate from the ideal tetrahedral angle of 109.47),[24-27] the configuration with the lowest energy of alternating spins inward and outward[11-16] is reached. Statistically, half of the molecules should be in state A and the other half in state B (Fig. 1e). It remains to be explored if whether the populations of the two states can be manipulated so information can be stored in $T'_1 > T > T_4$.

The measured dielectric constant is along the direction normal to the molecular planes, it is comprised of a small dipole moment $\varepsilon_z$ (arising from the slight tilting of the methyl groups) and the $\varepsilon_{zz}$ component of the quadrupole moment tensor. The region $T_1 > T > T_2$ is likely where the methyl groups elongate as this decreases both $\varepsilon_z$ and $\varepsilon_{zz}$. Then, the further out-of-plane tilting of the methyl groups occur in $T_2 > T > T_3$ since this increases $\varepsilon_z$. This triggers the system to phase transition into the near-cubic state in the region $T_3 > T > T_4$ which involves a shearing process between molecular layers and partial rotation of the methyl groups.[4]

As the system is warmed back up past $T'_4$, the methyl groups disorder; this renders the molecules to restore their previous phase **II** symmetry which triggers again the shearing process of the molecular layers and partial methyl group rotation lasting up to $T'_1$ and rendering the system to phase transition from **III** to **II**. No intermediate phases (Figs. 1b and 1d) exist because there is no energy degeneracy problem upon disordering. However, as the spins disorder, the resulting configurations do not fall far away from the ordered state configuration. This explains the "memory effects" observed in the specific heat reported in Ref. 28. The λ-peak present in the specific heat of HMB is well documented,[28-31] however, this is only present during warming from **III** to **II**. In contrast to when the system is cooled from **II** to **III**, the λ-peak is no longer present and only a small hump is observed. To this day, such metastable behavior remains unexplained. Moreover, apart from our recent work,[10] no claims have been made of the λ-peak being related to magnetic ordering despite that λ-peaks are usually associated to order-disorder phenomena such as onset of ferromagnetism & antiferromagnetism, spin order reorientations and onset of superconductivity & superfluidity. Furthermore, the specific heat usually only exhibits a step-like or a change-of-slope type of discontinuity during structural phase transitions. From such, we interpret the hump anomaly in the specific heat during cooling to be due to the system attempting to resolve the spatial energy degeneracy problem and the λ-peak due to a simple order to disorder transition of the spins of the methyl groups.

In contrast to Jahn-Teller distortions based on a spatial degeneracy of electronic ground states, the distortions in HMB occur at a slower pace in the sense that the process spans a larger temperature range which may be due to the following two reasons: a) a larger amount of energy is needed to realign the directions of the orbital motions of the much more massive protons, and/or b) a larger energy difference is needed to search for a minimum of a 20-fold energy degenerate system. From such, it is deduced that magnetic interactions stemming from orbital motions of protons should also play a key role in the structural stability of many other molecular solids. For instance, phase III of "the simplest organic molecule", solid methane, possesses a unique orthorhombic structure below 22.1 K.[32] At temperatures above, a fraction of the CH$_4$ rotors become ordered.[33] Another example is ammonium iodide where above and below $T_{\text{II-III}}$=235 K the ammonium cations are disordered and ordered in an anti-parallel array, respectively.[34] In phase III below $T_{\text{II-III}}$, the tetragonal structure is also one that is slightly distorted with the iodine ions oppositely displaced by around 0.012 Å along the plane of the nearest H atoms.[35] Preliminary measurements of the magnetic susceptibility show a pronounced increase at $T_{\text{II-III}}$ (not yet published) indicative of the geometric distortions likely arising from magnetic interactions between rotating NH$_4^+$ ions. Interestingly, dielectric constant measurements also reveal a large hysteretic region circa $T_{\text{II-III}}$.[36]

**Conclusions:**

In conclusion, the spin configurations of the methyl groups in triclinic hexamethylbenzene **II** are found to possess a 20-fold ground state degeneracy when cooled below 118 K. To lift the energy degeneracy, the molecules lower their symmetry according to the Jahn-Teller effect by readjusting the geometry of its methyl groups. The contortions of the methyl groups ultimately trigger a structural phase transition into the near-cubic phase **III**. The metastable nature of the λ-peak in the specific heat is explained to stem from the notion that the methyl groups require more energy to order than to disorder. From such, akin to phase III of solid methane,[33] it should be impossible to directly grow single crystals of HMB **III** as it is only accessible via phase **II**. The near-static dielectric constant has not been measured for most molecular solids where ordering of the periodic rotational motion of protons occur. We encouraged experimentalists to re-investigate the hundreds if not thousands of such compounds[37,38] as they may play a role in determining the structural properties at low temperatures and slightly elevated pressures. Lastly, we suggest modelers to incorporate this new driving mechanism into their crystal structure prediction packages.

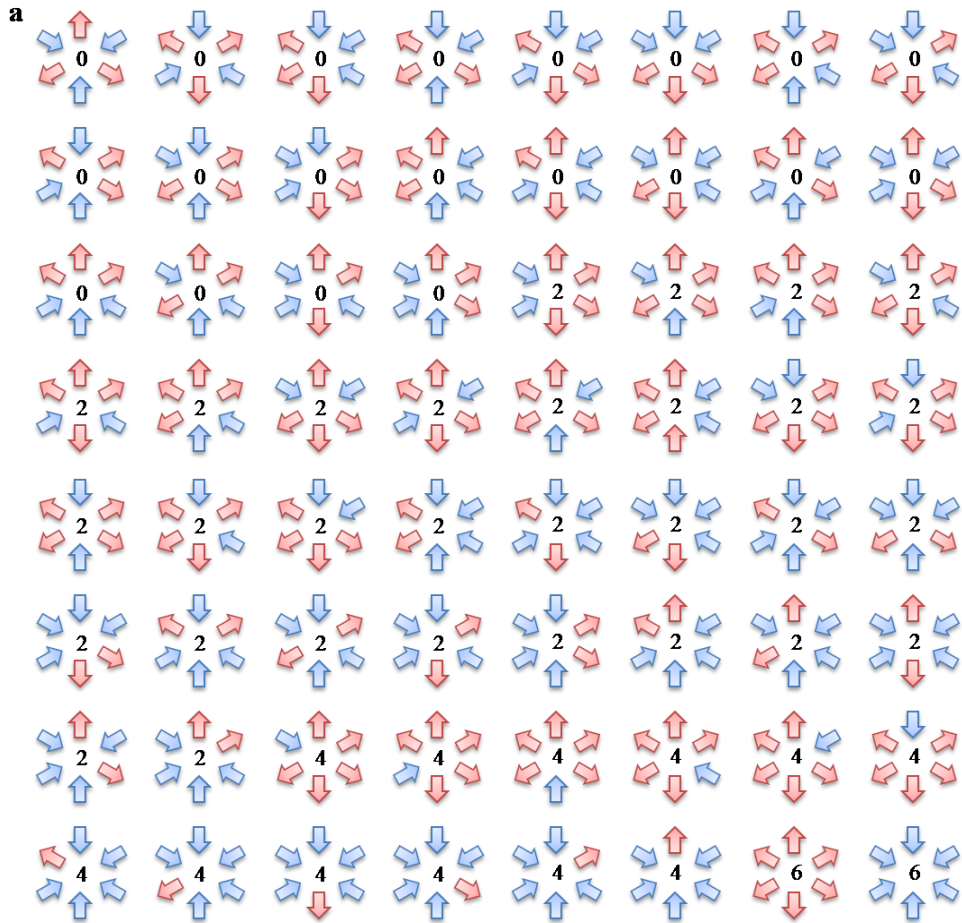

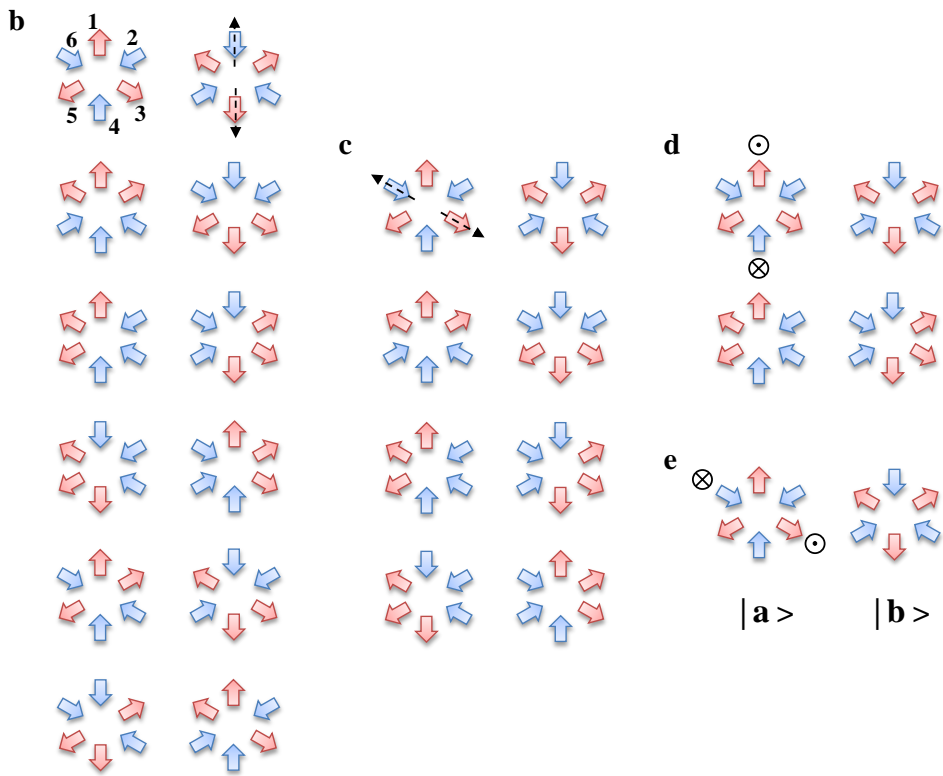

**Figure 1: a.** All $2^6=64$ possible configurations the methyl groups can rotate either along the clockwise (blue arrows) or counterclockwise (red arrows) directions in the triclinic phase **II** of hexamethylbenzene (HMB). The carbon atoms are not shown for clarity. The numbers in the middle are their respective energy values derived from the absolute value of the summation of blue (+1) and red (–1) arrows. The ground state is 20-fold degenerate; the next exited state has a 30-fold degeneracy, followed by 12-fold and 2-fold degeneracies. To lift the spatial energy degeneracy, the methyl groups elongate (dashed arrows) to three sets of different values. **b.** The methyl groups at sites *1* and *4* are displaced outward invoking the pair to have to possess opposing spins which reduces the ground state degeneracy to 12-fold **c.** The methyl groups *3* and *6* are also displaced outward but to a different value which reduces the degeneracy to 8-fold. **d.** The methyl groups at sites *1* and *4* tilt further out-of plane rendering its nearest adjacent methyl groups *2* and *5* to spin in the opposite direction. **e.** The methyl group sites *3* and *6* tilt to a different set of angles. This results in the methyl groups to alternatively rotate clockwise and anticlockwise (either phase $|a>$ or $|b>$) in each molecule in the near-cubic phase **III** of HMB.

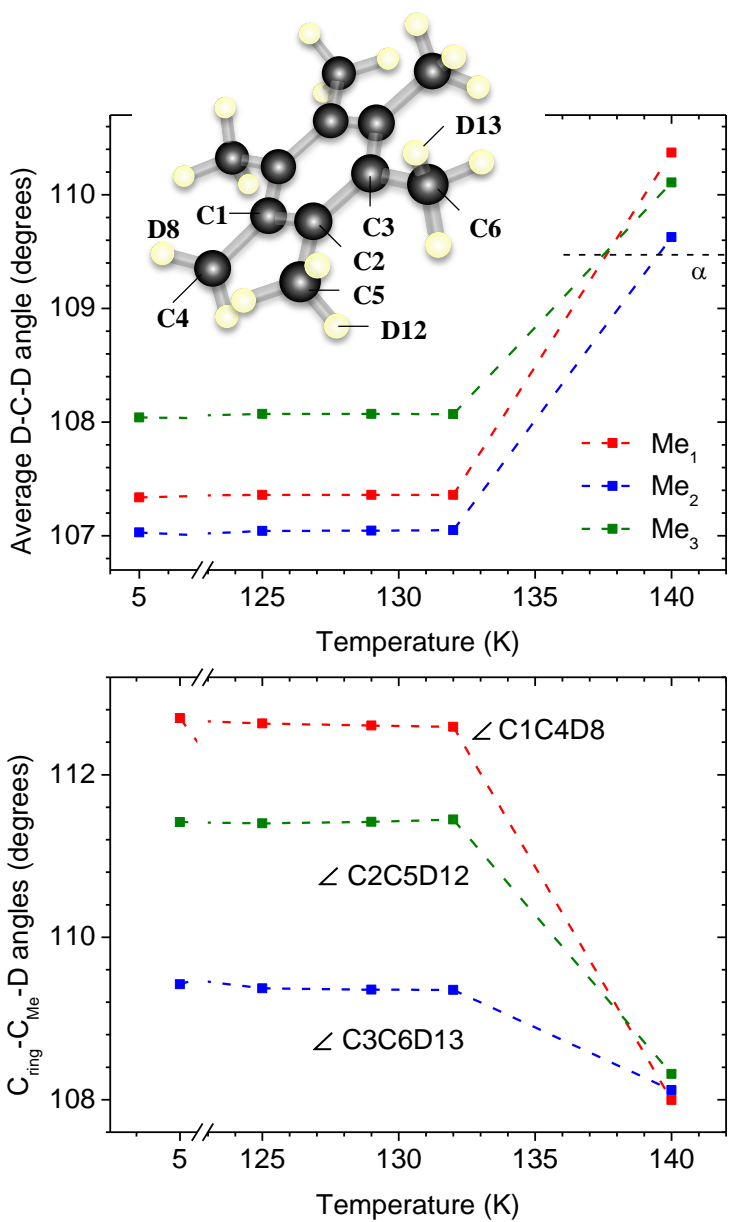

**Figure 2: a.** Average methyl group angles with respect to temperature according to neutron powder diffraction measurements on deuterated hexamethylbenzene (HMB-$d_{18}$) in the **II** and **III** phases;[4] $\alpha=109.47$ is the ideal tetrahedral angle for reference. Note that $T'_1$ for HMB-$d_{18}$ occurs near 132 K. Inset is the labeled HMB molecule. **b.** Out-of-plane tilting in the **II** and **III** phases.

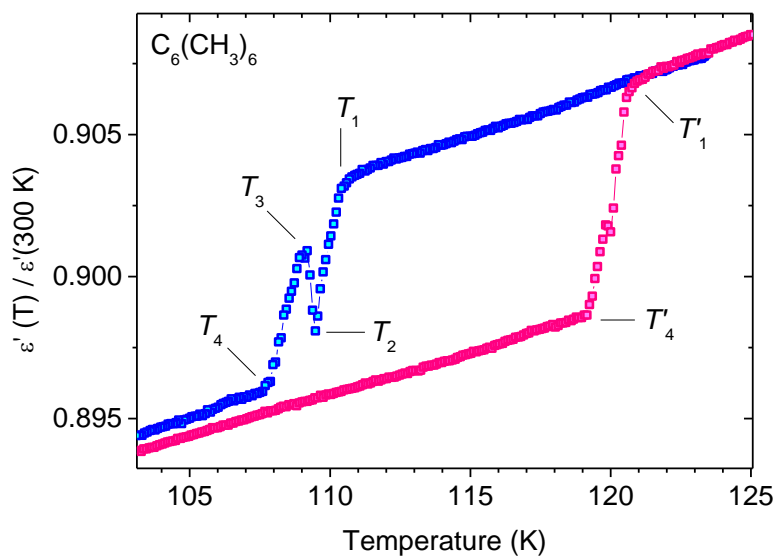

**Figure 3:** Dielectric constant of HMB measured at 10 kHz during cooling and warming. Upon cooling, at $T_1$=110.7 K, $T_2$=109.5, $T_3$=109.1 and $T_4$=107.8 the system undergoes a series of geometrical distortions described in the text in order to lift the spatial energy degeneracy of the methyl group rotations. During warming, the order to disorder transition (from Fig. 1e to 1a) occurs in nearly one step starting from $T'_4$=119.2 and ending at $T'_1$=120.9 K. In between, $T'_3$ and $T'_2$ are barely observable near 120.0 K.

**Acknowledgements:**

The author wishes to thank the National Natural Science Foundation of China for their support via grant numbers 11650110430 and 11374307.

**Table of Contents Graphic:**

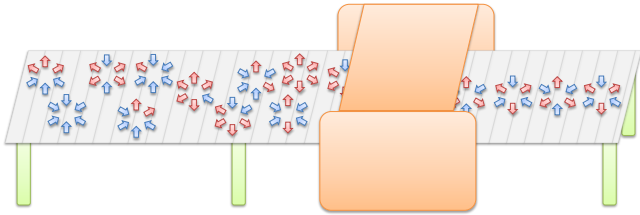